\documentstyle[aps,tighten,multicols]{revtex}
\begin{document}
%\preprint{EFEI-FIS}
\draft
\newcommand{\Fstar}{\raisebox{.2ex}{$\stackrel{*}{F}$}{}}
\newcommand{\astar}{\raisebox{-0.1ex}{$\stackrel{*}{a}$}{}}
\renewcommand{\thefootnote}{\fnsymbol{footnote}}
\title{Analog gravity from electrodynamics in non-linear media} 
\author{V. A. De Lorenci and R. Klippert } 
\address{ Instituto de Ci\^encias - 
Escola Federal de Engenharia de Itajub\'a \\
Av.\ BPS 1303 Pinheirinho, 37500-903 Itajub\'a, MG -- Brazil \\
(Electronic mail: {\tt lorenci@efei.br, klippert@efei.br})} 

\date{\today}
%\twocolumn[
%\hsize\textwidth\columnwidth\hsize\csname@twocolumnfalse\endcsname 
\maketitle

\begin{abstract}
\hfill{\small\bf Abstract}\hfill\smallskip
\par
Working with electrodynamics in the geometrical optics approximation 
we derive the expression representing an effectively curved geometry 
which guides the propagation of electromagnetic waves 
in material media whose physical properties depend on an 
external electric field.  
The issue of birefringence is addressed, 
and the trajectory of the extraordinary ray is explicitly worked out.  
Quite general curves are obtained for the path of the light ray 
by suitably setting an electric field.  
\end{abstract}
\pacs{PACS numbers: 04.20.-q, 11.10.-z}
\renewcommand{\thefootnote}{\arabic{footnote}}

\begin{multicols}{2}

\section{Introduction}

The theory of general relativity lead us to the
conception of a generally curved geometry
of the spacetime whose source encompass 
all matter and energy content of the universe. 
The comparative study between the kinematic aspects of
general relativity and other kinds of interactions has recently
been called as {\em analog models} for gravitation \cite{analog}. 
Indeed, it have extensively been examined, 
not only with respect to electrodynamics (non-linear
\cite{Plebanski,Heyl,Dittrich,Novello,DeLorenci,Liberati} 
as well as in moving dielectric fluids 
\cite{Leonhardt,LeonhardtPRL,Visser2,Brevik,Visser0}) 
but also for acoustic perturbations \cite{Unruh,Jacobson,Visser1}. 
Long ago, Gordon \cite{Gordon} shown that the refractive index 
of a medium can be reinterpreted by means of an effective metric 
$g^{\mu\nu}_{\scriptscriptstyle Gordon} = \eta^{\mu\nu} 
-  (1-\epsilon\mu)V^\mu V^\nu$, where $V_\mu$ represents the
velocity 4-vector of an arbitrary observer.  
The paths of light rays in non-linear electrodynamics can be obtained 
in terms of the geodesic equations of an effective geometry 
representing the non-linearities \cite{Dittrich,Novello}. 
These works assumed electrodynamics 
as governed by a quite general non-linear Lagrangian 
which is a function of the two invariants of the Maxwell field.  

Inside material media, Maxwell equations 
must be supplemented by constitutive laws that relate the electromagnetic 
excitations and the field strengths by means of quantities characterizing
each medium the waves propagate into.  
In this context, the electromagnetic interaction was geometrized 
for the case of linear constitutive relations \cite{Obukhov,Rubilar}.
More recently, specific non-linear constitutive relations 
were also geometrized \cite{Barbati,Souza}.  
The above results lead us to conclude that electromagnetic waves 
propagate inside material media as if immersed in a curved spacetime.  
This fact allows one to make an analogy between wave 
propagation in non-linear media and gravitational phenomena.  

This work explicitly presents, in the context of geometrical optics, 
the construction of an effective geometry which describes 
small perturbations of the electromagnetic field inside substances 
whose electric susceptibility depend on the electric field.  
A description of birefringence \cite{Souza} observed for this class 
of materials is presented, and the problem of the trajectory 
of the light ray is explicitly presented in a similar way 
as the bending of light due to gravitational interaction.  

We work in Minkowski spacetime employing a Cartesian coordinate system.
The background metric will be represented by $\eta_{\mu\nu}$,
which is defined by $diag(+1,-1,-1,-1)$. 
Units are such that $c=1$.

\section{Field equations}

$F^{\mu\nu}$ and  $P^{\mu\nu}$ are the tensors 
representing the total electromagnetic field, which are
expressed in terms of the strengths and 
the excitations of the electric and magnetic fields 
as
\begin{eqnarray}
F^{\mu\nu} &=& V^{\mu}E^{\nu} - V^{\nu}E^{\mu} 
- \eta^{\mu\nu}{}_{\alpha\beta}V^{\alpha}B^{\beta}
\label{1}
\\
P^{\mu\nu} &=& V^{\mu}D^{\nu} - V^{\nu}D^{\mu} 
- \eta^{\mu\nu}{}_{\alpha\beta}V^{\alpha}H^{\beta}
\label{2}
\end{eqnarray}
where the Levi-Civita tensor
introduced is defined such that $\eta^{0123} = +1$.

In general the properties of the media 
are determined by the tensors 
$\varepsilon^\alpha{}_\beta$ and $\mu^\alpha{}_\beta$ which
relate the electromagnetic excitation and the field strength by
the generalized  constitutive laws,
\begin{eqnarray}
D^{\alpha} &=&  \varepsilon^{\alpha}{}_{\beta}(E^\mu,\,H^\mu)E^{\beta}
\label{8}
\\ 
B^{\alpha} &=& \mu^{\alpha}{}_{\beta}(E^\mu,\,H^\mu)H^{\beta}.
\label{9}
\end{eqnarray} 

In the absence of sources Maxwell\rq s 
theory can be summarized by the equations
\begin{eqnarray}
P^{\mu\nu}{}_{,\nu} &=& 0
\label{13}
\\
\Fstar^{\mu\nu}{}_{,\nu} &=& 0
\label{14}
\end{eqnarray} 
together with the constitutive laws (\ref{8}) and (\ref{9}), 
where we have introduced the dual electromagnetic field tensor
\begin{equation}
\Fstar^{\mu\nu} = \frac{1}{2}\eta^{\mu\nu}{}_{\alpha\beta}
F^{\alpha\beta}.
\label{11}
\end{equation}
Eq.\ (\ref{14}) could be alternatively obtained by the Bianchi identity.
Now, rewriting the above field equations in terms of the
electric and magnetic fields it results
\begin{eqnarray}
V^{\mu}D^{\nu}{}_{,\nu} - V^{\nu}D^{\mu}{}_{,\nu} 
- \eta^{\mu\nu\alpha\beta}V_{\alpha}H_{\beta,\nu}&=&0
\label{16}
\\
V^{\mu}B^{\nu}{}_{,\nu} - V^{\nu}B^{\mu}{}_{,\nu} 
+ \eta^{\mu\nu\alpha\beta}V_{\alpha}E_{\beta,\nu}&=&0.
\label{18}
\end{eqnarray} 

Additionally, the electromagnetic excitation is related to the
field strength by means of the constitutive relations (\ref{8}) and
(\ref{9}), whose derivatives with respect to the coordinates
can be presented as
\begin{eqnarray}
D^\alpha{}_{,\tau} &=& \varepsilon^{\alpha}{}_{\beta}E^\beta{}_{,\tau} 
+ \frac{\partial \varepsilon^\alpha{}_{\beta}}{\partial
E^\mu}E^{\beta}E^\mu{}_{,\tau} + \frac{\partial       
\varepsilon^{\alpha}{}_{\beta}}{\partial H^\mu}E^{\beta}H^\mu{}_{,\tau}
\label{28} 
\\
B^\alpha{}_{,\tau} &=& \mu^{\alpha}{}_{\beta}H^\beta{}_{,\tau}
+\frac{\partial \mu^{\alpha}{}_{\beta}}{\partial E^\mu}
H^{\beta}E^\mu{}_{,\tau} 
+\frac{\partial \mu^{\alpha}{}_{\beta}}{\partial
H^\mu}H^{\beta}H^\mu{}_{,\tau}.
\label{29} 
\end{eqnarray}

\section{Geometrical optics}

In order to determine the propagation of waves associated
to the electromagnetic field, we will consider the 
method of field discontinuities \cite{Hadamard}. 
We define a surface of discontinuity $\Sigma$ by
$z(x^{\mu}) = 0$. Whenever $\Sigma$ is a global surface, it 
divides the spacetime  in two distinct regions $U^-$ for $z(x^\mu)<0$,
and $U^+$ for $z(x^\mu)>0$.  The discontinuity of an arbitrary 
function $f(x^\alpha)$ on $\Sigma$ is given by
\begin{equation}
\label{discontinuity}
\left[f(x^{\alpha})\right]_{\Sigma} 
\doteq \lim_{\{P^\pm\}\rightarrow P}
\left[f(P^+) - f(P^-)\right]
\end{equation}
with $P^+,\,P^-$ and $P$ belonging to $U^+,\,U^-$ 
and $\Sigma$, respectively.
The electric and magnetic fields are continuous when crossing
the surface $\Sigma$.  
However, their derivatives behave as 
\begin{equation}
\left[E^\mu{}_{,\nu}\right]_{\Sigma} = e^{\mu}K_{\nu};
\;\;\;\;
\left[H^\mu{}_{,\nu}\right]_{\Sigma} = h^{\mu}K_{\nu}
\label{35}
\end{equation}
where $e^{\mu}$ and $h^{\mu}$ represent the discontinuities of 
the fields on the surface $\Sigma$ and 
\begin{equation}
\label{K}
K_{\lambda}=\frac{\partial\Sigma}{\partial x^\lambda}
\end{equation} 
is the wave vector.
Applying these conditions to the field equations (\ref{16}) and 
(\ref{18}), we obtain 
\end{multicols}
\vspace{0.2cm}
\ruleleft
\vspace{0.2cm} 
\baselineskip=13pt
\begin{eqnarray}
\varepsilon^\alpha{}_\beta K_{\alpha}e^{\beta} 
+\frac{\partial \varepsilon^\alpha{}_\beta}{\partial   
E^\mu}E^{\beta}K_{\alpha}e^{\mu} +\frac{\partial
\varepsilon^\alpha{}_\beta}{\partial H^\mu}E^{\beta}
K_{\alpha}h^{\mu} &=& 0
\label{224} 
\\ 
\mu^\alpha{}_\beta K_{\alpha}h^{\beta} 
+\frac{\partial \mu^\alpha{}_\beta}{\partial E^\mu}
H^{\beta}K_{\alpha}e^{\mu} +\frac{\partial
\mu^\alpha{}_\beta}{\partial H^\mu}
H^{\beta}K_{\alpha}h^{\mu} &=& 0
\label{226} 
\\
\left(\varepsilon^\mu{}_\beta e^{\beta} 
+\frac{\partial \varepsilon^\mu{}_\beta}{\partial   
E^\alpha}E^{\beta}e^{\alpha} +\frac{\partial     
\varepsilon^\mu{}_\beta}{\partial   
H^\alpha}E^{\beta}h^{\alpha}\right)(KV) +
\eta^{\mu\nu}{}_{\alpha\beta}K_{\nu}V^{\alpha}h^{\beta} &=& 0  
\label{225} 
\\
\left(\mu^\mu{}_{\beta}h^{\beta} 
+\frac{\partial \mu^\mu{}_\beta}{\partial  
E^\alpha}H^{\beta}e^{\alpha} 
+\frac{\partial      
\mu^\mu{}_\beta}{\partial H^\alpha}H^{\beta}h^{\alpha}
\right)(KV) -
\eta^{\mu\nu}{}_{\alpha\beta}K_{\nu}V^{\alpha}e^{\beta} &=& 0
\label{227}
\end{eqnarray}
\vspace{0.2cm}
\ruleright
\begin{multicols}{2}
\baselineskip=12pt
\noindent
where we have defined $(KV) \doteq K^{\mu}V_{\mu}$.
Eqs.\ (\ref{224})--(\ref{226}) 
came from the zeroth component of Eqs.\ (\ref{16})--(\ref{18}), 
and can be obtained from Eqs.\ (\ref{225})--(\ref{227}) 
upon contraction with $K_\mu$ whenever $(KV)\neq0$.  
Then, it follows that the coupled system (\ref{225})--(\ref{227}) 
completely describe the propagation of light rays.  

For the cases 
$\varepsilon_{\alpha\beta} = \varepsilon_{\alpha\beta}(E^\mu,H^\mu)$
and $\mu_{\alpha\beta} = \mu(\eta_{\alpha\beta}-V^\alpha V^\beta)$ 
with $\mu$ a constant, Eqs.\ (\ref{225})--(\ref{227}) reduces to 
\begin{equation}
\label{h}
h^\alpha=
\frac{1}{\mu(KV)}\eta^{\alpha\beta\sigma\rho}K_\beta V_\sigma e_\rho
\end{equation}
and the eigen-vector equation
\end{multicols}
\vspace{0.2cm}
\ruleleft
\vspace{0.2cm} 
\baselineskip=13pt
\begin{equation}
\left[ \mu (KV)^2\!\!\left(\!\varepsilon^\mu{}_\beta
+\frac{\partial \varepsilon^\mu{}_\alpha}{\partial   
E^\beta}E^{\alpha}\!\!\right)
+(KV)\frac{\partial      
\varepsilon^\mu{}_\rho}{\partial H^\alpha}\eta^{\alpha\tau\sigma}{}_\beta
E^{\rho}K_{\tau}V_{\sigma}
-K^{\mu}K_{\beta}+(KV)V^{\mu}K_{\beta}
+K^{2}\delta^\mu_\beta-(KV)^{2}\delta^\mu_\beta\right]
e^{\beta} = 0
\label{235} 
\end{equation}
\vspace{0.2cm}
\ruleright
\begin{multicols}{2}
\baselineskip=12pt
\noindent
The Fresnel equation represents 
non-trivial solutions of Eq.\ (\ref{235}) 
and is given by the determinant of the term in brackets.  
The solution of the Fresnel equation for linear constitutive 
relations is derived in reference \cite{Rubilar}.

Now, let us analyze non-linear media whose physical
properties are influenced by an external electric field as 
\begin{equation}
\varepsilon^{\mu\beta} = 
\epsilon(E^\lambda)(\eta^{\mu\beta}-V^\mu V^\beta) - 
\alpha(E^\lambda) E^{\mu}E^{\beta}.
\label{240}
\end{equation}
Each particular medium is characterized by the
parameters $\alpha$ and $\epsilon$.
Considering Eq.\ (\ref{240}), 
the eigen-vector problem reduces to
\end{multicols}
\vspace{0.2cm}
\ruleleft
\vspace{0.2cm} 
\baselineskip=13pt
\begin{equation}
\left\{\left[K^2-(KV)^2 + \mu(KV)^2(\epsilon+\alpha E^2)\right]
\delta^\mu_\beta-K^\mu K_\beta +(KV)V^\mu V_\beta 
-E^\mu \left(2\alpha E_\beta
-\frac{\partial\epsilon}{\partial E^\beta} 
-E^2\frac{\partial\alpha}{\partial E^\beta}\right)
\right\} e^{\beta} = 0
\label{a4} 
\end{equation}
where we denoted $E^2=-E^{\alpha}E_{\alpha}$. 
Expanding the polarization vector in a suitable basis 
\begin{equation}
e^\beta=aE^\beta+bH^\beta+cK^\beta+dV^\beta
\end{equation} 
one gets 
\begin{eqnarray}
&&a\left[\frac{K^2}{\mu(KV)^2}-\frac{1}{\mu}
+\epsilon+3\alpha E^2
+E^\beta\left(\frac{\partial\epsilon}{\partial E^\beta}
+E^2\frac{\partial\alpha}{\partial E^\beta}\right)\right]
+b\left[-2\alpha E^\beta H_\beta
+H^\beta\left(\frac{\partial\epsilon}{\partial E^\beta}
+E^2\frac{\partial\alpha}{\partial E^\beta}\right)\right]
\nonumber\\
&&+c\left[-2\alpha E^\beta K_\beta
+K^\beta\left(\frac{\partial\epsilon}{\partial E^\beta}
+E^2\frac{\partial\alpha}{\partial E^\beta}\right)
\right]=0
\label{a-10}\\
&&b\left[K^2-(KV)^2+\mu(KV)^2(\epsilon+\alpha E^2)\right]=0
\label{a-11}\\
&&aE^\mu K_\mu +bH^\mu K_\mu 
+c(KV)^2\left[1-\mu(\epsilon+\alpha E^2)\right]+d(KV)=0
\label{a-12}\\
&&a(KV)E^\mu K_\mu +b(KV)H^\mu K_\mu 
+c(KV)K^2+d\left[K^2+\mu(KV)^2(\epsilon+\alpha E^2)\right]=0.
\label{a-13}
\end{eqnarray}
\vspace{0.2cm}
\ruleright
\begin{multicols}{2}
\baselineskip=12pt
\noindent
There are two distinct solutions of the above system. First,
by fixing $b\neq0$, it follows that $a=c=d=0$, and we obtain the
light cone condition
\begin{equation}
K^2=(KV)^2\left[1-\mu\frac{\Omega}{E}\right]
\label{b-9}
\end{equation}
where $\Omega\equiv E(\epsilon+\alpha E^2)$. Conversely, assuming
$b=0$ and $a \neq 0$, it follows that $c= [E\,E^\beta K_\beta/
\mu\Omega(KV)^2]\,a$ and  $\qquad d=-[E\,E^\beta
K_\beta/\mu\Omega(KV) ]\,a$, and results in the light cone condition
\begin{equation}
K^2\!=\!(KV)^2\!\!\left(\!\!1\!-\!\mu\frac{E^\beta}{E}
\frac{\partial\Omega}{\partial E^\beta}\!\!\right)\!
-\!\frac{E E^\alpha K_\alpha K^\beta}{\Omega}\frac{\partial}{\partial E^\beta}
\!\left(\!\frac{\Omega}{E}\!\right)
\label{b-1}
\end{equation}
Therefore, there are different light-cone conditions 
for each polarization eigenvectors $e^\beta$.  
Corresponding to Eq.\ (\ref{b-9}) one has the polarization mode 

\begin{equation}
e^{\beta}_{\scriptscriptstyle+}=b\,H^\beta
\label{e+}
\end{equation}
traveling with phase velocity 
\begin{equation}
\label{v+}
v_{\scriptscriptstyle+}^2=\frac{E}{\mu\Omega}.
\end{equation}
Moreover, Eqs.\ (\ref{a-10}), (\ref{a-12})--(\ref{a-13}) 
for this mode require the following conditions: 
$H^\beta K_\beta=0$, 
$\alpha(E^\lambda H_\lambda)(E^\beta K_\beta)=0$,  and
$H^\beta(\partial/\partial E^\beta)(\Omega/E)=0$.  
Similarly, from Eq.\ (\ref{b-1}) the polarization mode and its phase
velocity are given by 
\begin{eqnarray}
\label{e-}
e^{\beta}_{\scriptscriptstyle-}&=&c\left[\frac{\mu(KV)^2}{E^\lambda
K_\lambda}\,\frac{\Omega}{E}E^\beta+K^\beta-(KV)V^\beta \right]\\
\label{v-}
v_{\scriptscriptstyle-}^2&=&\frac{E}{\mu E^\beta
\frac{\textstyle\partial\Omega}{\textstyle\partial E^\beta}}\left[1-
\frac{E}{\Omega}E^\gamma\hat{k}{}_\gamma
\hat{k}{}^\alpha\frac{\partial}{\partial E^\alpha}
\left(\frac{\Omega}{E}\right)\right]
\end{eqnarray}
where we introduced the space-like unit propagation vector 
$\hat{k}{}^\alpha=(\eta^{\alpha\beta}-V^\alpha
V^\beta)K_\beta/\sqrt{(KV)^2-K^2}$.   
Contrarily to the velocity $v_{\scriptscriptstyle+}$, 
the velocity $v_{\scriptscriptstyle-}$ 
depends on the direction of the wave propagation.  
Both these velocities coincide when the wave propagates 
along the direction of the electric field
$E^\beta$ ({\em i.e.}, $\hat{k}{}^\beta=\pm E^\beta/E$).   
The appearance of two different modes with different velocities 
points out to birefringence phenomena \cite{Landau,Born}, 
in this case induced by the electric field. 
The modes (\ref{e+}) and (\ref{e-}) correspond to 
the ordinary and the extraordinary rays, respectively.  
The corresponding refraction indexes are given by $n_\pm=1/v_\pm$. 

\section{Analog gravity}
Eqs.\ (\ref{b-9})--(\ref{b-1}) 
can be restated in the more appealing form  
\begin{equation}
g^{\mu\nu}K_\mu K_\nu=0
\label{g1}
\end{equation}
where we defined the symmetric tensors 
\begin{eqnarray}
g^{\mu\nu}_+&=&\eta^{\mu\nu}-
\left(1-\mu\frac{\Omega}{E}\right)V^\mu V^\nu
\label{g+}\\
g^{\mu\nu}_-&=&\eta^{\mu\nu}-\left(1-\mu\frac{E^\beta}{E}
\frac{\partial\Omega}{\partial E^\beta}\right)V^\mu V^\nu
\nonumber\\
&&+\frac{E}{2\Omega}\frac{\partial}{\partial E^\beta}
\left(\frac{\Omega}{E}\right)\eta^{\beta(\mu}E^{\nu)}
\label{g-}
\end{eqnarray}
in which $a^{(\mu\nu)}\equiv a^{\mu\nu}+a^{\nu\mu}$ 
for an arbitrary tensor $a^{\mu\nu}$. 
The inverse symmetric tensor $g_{\mu\nu}$ is defined in such way that 
$g^{\mu\alpha}g_{\alpha\nu}=\delta^\mu_\nu$.  

Differentiation of Eq.\ (\ref{g1}) with respect to $x^\lambda$ yields
\begin{equation}
\label{N16}
2K_{\nu,\,\lambda}K_\mu g^{\mu\nu} +
K_\mu K_\nu g^{\mu\nu}\mbox{}_{,\,\lambda} = 0.
\end{equation}
As the vector $K_\mu$ is a gradient (\ref{K}) it follows that 
$K_{\mu,\,\lambda}=K_{\lambda,\,\mu}$. 
Moreover, one has the identity 
\begin{equation}
g^{\mu\nu}{}_{,\,\lambda}=-
g^{\nu\sigma}g^{\mu\alpha}(g_{\alpha\sigma,\,\lambda}+
g_{\alpha\lambda,\,\sigma}-g_{\sigma\lambda,\,\alpha}).
\end{equation}
Thus, Eq.\ (\ref{N16}) can be presented as 
\begin{equation}
\label{geodesic}
g^{\mu\nu}K_\mu\left[K_{\lambda,\,\nu}- 
\Gamma^\alpha{}_{\nu\lambda}K_\alpha\right]=0
\end{equation}
where we defined the quantity 
\begin{equation}
\label{Christoffel}
\Gamma^\mu{}_{\sigma\lambda}=\frac{1}{2}
g^{\mu\alpha}(g_{\alpha\sigma,\,\lambda}+
g_{\alpha\lambda,\,\sigma}-g_{\sigma\lambda,\,\alpha}).
\end{equation}

Eqs.\ (\ref{g+}) and (\ref{g-}) depend on the total field 
$\vec{E}$, which encompasses both the external field 
and the one associated with the wave itself.  
Thus, the standard treatment of birefringence \cite{Landau,Born} 
by setting a frequency-dependent $\epsilon^\alpha{}_\beta(\omega)$ 
is included in the present formalism as a particular case as 
$\partial/\partial\omega=
(\partial E^\nu/\partial\omega)\partial/\partial E^\nu$.  
We will hereafter adopt the point of view that 
the wave fields are much smaller than their external counterparts.  
Thus, the fields appearing in $g^{\mu\nu}$ 
are independent of the light ray.  

Therefore one recognizes that, 
for the propagation vector $K_\nu$, 
the tensor $g^{\mu\nu}$ is a metric tensor.  
Indeed, $K_\nu$ is a light-like (or null) vector 
with respect to this metric tensor, Eq.\ (\ref{g1}), 
and also satisfies the geodesic equation (\ref{geodesic}) 
in terms of the Christoffel symbols $\Gamma^\alpha{}_{\!\mu\nu}$ 
associated with this metric, Eq.\ (\ref{Christoffel}).  
We thus refer to $K_\nu$ as a geodesic null vector 
with respect to the effective metric tensor $g^{\mu\nu}$.  
The two terms in brackets in Eq.\ (\ref{geodesic}) 
correspond to the covariant derivative of $K_\nu$ 
and will henceforth be denoted as $K_{\nu;\,\lambda}$.  
The effective geometry $g^{\mu\nu}$ allows us to write 
$K^\mu=g^{\mu\nu}K_\nu$, 
for which the geodesic equation can be written in the simpler form 
\begin{equation}
\label{geodesic2}
K^\mu{}_{;\,\lambda}K^\lambda=0
\end{equation}
which avoids explicit dependence on the metric tensor.  

\section{Trajectories}

The trajectories of light-rays are the geodesic lines 
which solve the geodesic equation (\ref{geodesic}).  
For the sake of simplicity, 
we will consider $V^\nu=\delta^\nu_o$.  
Only the calculations for the extraordinary ray will be presented, 
as the corresponding ones for the ordinary ray 
follow exactly the same reasoning (being much simpler, however).  
In order to fit it best problems with axial symmetry, 
standard cylindrical coordinates $x^\mu=(t,\,r,\,\theta,\,z)$ 
will be adopted.  
For a configuration of electric field $E^\mu=(0,\,\vec{E})$ with 
$\vec{E}=(E_r,\,E_\theta,\,E_z)=E_z(r,\,\theta)\hat{z}$, and assuming 
$\partial\Omega/\partial E_r=0=\partial\Omega/\partial E_\theta$, 
the metric associated with the extraordinary ray reduces to 
\begin{equation}
\label{g-EH}
g^{\mu\nu}=\pmatrix{\mu\frac{E_z}{E}\frac{\partial\Omega}{\partial E_z}
  & 0  & 0              & 0 \cr
0 & -1 & 0              & 0 \cr
0 & 0  & -\frac{1}{r^2} & 0 \cr
0 & 0  & 0              & 
-\frac{E_z}{\Omega}\frac{\partial\Omega}{\partial E_z}}.
\end{equation}
Thus, 
$g_{\mu\nu}={diag}[f_1(r,\,\theta),\,-1,\,-r^2,\,-f_2(r,\,\theta)]$, 
where we introduce the notation 
\begin{eqnarray}
f_1(r,\,\theta)&=&\left(\mu\frac{E_z}{E}
\frac{\partial\Omega}{\partial E_z}\right)^{-1} \\
f_2(r,\,\theta)&=&\left(\frac{E_z}{\Omega}
\frac{\partial\Omega}{\partial E_z}\right)^{-1}.
\end{eqnarray}
We will also denote the components of the propagation vector 
as $K^\nu=(dt,\,dr,\,d\theta,\,dz)/d\tau$, 
where $\tau$ is an affine parameter along the integral curves 
of the vector $K^\nu$, and $(t,\,r,\,\theta,\,z)$ 
represent the coordinates of the points belonging to these lines.  
The geodesic lines for this case are given by 
Eq.\ (\ref{geodesic2}) as 
\begin{eqnarray}
\label{T}
\frac{d^2t}{d\tau^2}&=&-\frac{1}{f_1}\frac{\partial f_1}{\partial x^\beta}
\frac{dx^\beta}{d\tau}\frac{dt}{d\tau} \\
\label{R}
\frac{d^2r}{d\tau^2}&=&\frac{1}{2}\left[\frac{\partial f_2}{\partial r}
\left(\frac{\partial z}{\partial\tau}\right)^2
-\frac{\partial f_1}{\partial r}
\left(\frac{\partial t}{\partial\tau}\right)^2\right]
+r\left(\frac{d\theta}{d\tau}\right)^2 \\
\label{TH}
\frac{d^2\theta}{d\tau^2}&=&
\frac{1}{2r^2}\left[\frac{\partial f_2}{\partial \theta}
\left(\frac{\partial z}{\partial\tau}\right)^2
-\frac{\partial f_1}{\partial\theta}
\left(\frac{\partial t}{\partial\tau}\right)^2\right]
-\frac{2}{r}\frac{dr}{d\tau}\frac{d\theta}{d\tau} \\
\label{Z}
\frac{d^2z}{d\tau^2}&=&-\frac{1}{f_2}\frac{\partial f_2}{\partial x^\beta}
\frac{dx^\beta}{d\tau}\frac{dz}{d\tau}.
\end{eqnarray}
The solutions of Eqs.\ (\ref{T}) and (\ref{Z}) are 
\begin{eqnarray}
\label{tdot}
\frac{dt}{d\tau}&=&\frac{A}{f_1} \\
\label{zdot}
\frac{dz}{d\tau}&=&\frac{B}{f_2}
\end{eqnarray}
where $A$ and $B$ are arbitrary constants.  
Inserting these relations in  Eqs.\ (\ref{R})--(\ref{TH}) one gets 
\begin{eqnarray}
\label{ddR}
\frac{d^2r}{d\tau^2}&-&r\left(\frac{d\theta}{d\tau}\right)^2=
\frac{1}{2}\frac{\partial}{\partial r}
\left[\frac{A^2}{f_1}-\frac{B^2}{f_2}\right] \\
\label{ddTheta}
\frac{d^2\theta}{d\tau^2}
&+&\frac{2}{r}\frac{dr}{d\tau}\frac{d\theta}{d\tau}=
\frac{1}{2r^2}\frac{\partial}{\partial\theta} 
\left[\frac{A^2}{f_1}-\frac{B^2}{f_2}\right].
\end{eqnarray}
For $dr/d\tau\neq0\neq d\theta/d\tau$ 
equations (\ref{ddR})--(\ref{ddTheta}) lead to 
\begin{equation}
\label{D}
\left(\frac{dr}{d\tau}\right)^2+r^2\left(\frac{d\theta}{d\tau}\right)^2
=\frac{A^2}{f_1}-\frac{B^2}{f_2}+D
\end{equation}
where $D$ is an arbitrary constant.  
Introducing Eqs.\ (\ref{tdot})--(\ref{zdot}) and (\ref{D}) 
in the light-cone constraint 
\begin{equation}
\label{ds2}
ds^2=g_{\mu\nu}dx^\mu dx^\nu=g_{\mu\nu}K^\mu K^\nu d\tau^2=0
\end{equation}
it follows that $D=0$.  Thus, 
Eq.\ (\ref{D}) yields $d\theta/d\tau$ in terms of $r$ which is given by 
\begin{equation}
\label{radial}
\frac{d^2(r^2)}{d\tau^2}=\frac{1}{r}\frac{\partial}{\partial r}
\left[r^2\left(\frac{A^2}{f_1}-\frac{B^2}{f_2}\right)\right].
\end{equation}

\subsection*{Cylindrical symmetry}

For the simple case $f_1=f_1(r)$ and $f_2=f_2(r)$ 
it follows from Eq.\ (\ref{ddTheta}) that 
\begin{equation}
\label{C}
\frac{d\theta}{d\tau}=\frac{C}{r^2}
\end{equation}
with an arbitrary constant $C$, from which Eq.\ (\ref{D}) yields 
\begin{equation}
\label{Rpm}
\frac{dr}{d\tau}=\pm\sqrt{\frac{A^2}{f_1}-\frac{B^2}{f_2}-\frac{C^2}{r^2}}.
\end{equation}
The trajectories of the light ray is then found from the ratios 
$(dz/d\tau)/(dr/d\tau)$ and $(d\theta/d\tau)/(dr/d\tau)$, 
from which one obtains $z=z(r)$ and $\theta=\theta(r)$ as 
\begin{eqnarray}
\label{zInt}
z&=&\pm B\int^r
\frac{dr}{f_2\sqrt{\frac{\textstyle A^2}{\textstyle f_1}
-\frac{\textstyle B^2}{\textstyle f_2}
-\frac{\textstyle C^2}{\textstyle r^2}}} \\
\label{thetaInt}
\theta&=&\pm C\int^r
\frac{dr}{r^2\sqrt{\frac{\textstyle A^2}{\textstyle f_1}
-\frac{\textstyle B^2}{\textstyle f_2}
-\frac{\textstyle C^2}{\textstyle r^2}}}.
\end{eqnarray}

Eqs.\ (\ref{tdot})--(\ref{zdot}), (\ref{D}) and (\ref{radial}) 
encompass in this way a complete description 
for the path of the light ray 
with arbitrary $\Omega(E_z)=E[\epsilon(E_z)+\alpha(E_z)E^2]$ 
as a function of $E_z(r,\,\theta)$ 
in terms of the arbitrary constants $A,\,B,\,C$.  
Thus, quite general curves can be obtained for the path of the light ray 
by suitably setting the electric field $E_z$ at the given medium.  
From $v^i=(dx^i/d\tau)/(dt/d\tau)$ it follows that 
\begin{equation}
v^2=
\frac{E}{\mu E_z\frac{\textstyle\partial\Omega}{\textstyle\partial E_z}}
+\left(\!\frac{B^2}{2\mu A^2}\!\right)^2\left[\!
\left(\!\frac{\Omega}{E_z}\!\right)^{-2}
\!\!-\!\left(\!\frac{\partial\Omega}{\partial E_z}\!
\right)^{-2}\right]
\end{equation}
which is to be compared with Eq.\ (\ref{v-}).  
Birefringence then occurs for this model whenever 
$|\partial\Omega/\partial E_x|\neq|\Omega/E_x|$, 
which can be ascribed either to a non zero $\alpha$ parameter 
or to a varying $\varepsilon(\vec{E})$ in equation (\ref{240}), 
or even to both of them.  

\section{Conclusion}

Maxwell linear equations together with 
the constitutive relations for non-linear materials 
lead to an effective non-linear theory of electrodynamics.  
The problem of the propagation of electromagnetic waves 
was here dealt with under the approximation of geometrical optics.  
The path of light rays thus belong to the characteristic surfaces 
which depend on both the external fields and the propagation fields.  
For the case of a constant scalar magnetic permeability 
a Fresnel problem is obtained \cite{Souza}.  
The propagating modes and their velocities were obtained 
for the class of electric permittivity $\varepsilon^{\mu\nu}(\vec{E})$.  
Weak waves were then described as light rays 
propagating along null geodesics of an effectively deformed geometry 
which depends on the external electric field, 
and birefringence phenomena were found to occur if 
$\partial\varepsilon^{\mu\nu}/\partial E^\lambda\neq0$.  
Explicit calculations for the path and velocity of the rays 
were presented for the particular 
$\varepsilon^{\mu\nu}=\epsilon(\vec{E})[\eta^{\mu\nu}-V^\mu V^\nu]
-\alpha(\vec{E})E^\mu E^\nu$ case for cylindrically symmetric media, 
illustrating the usefulness of the geometrical approach.  
The results of this investigation may be useful to test kinematic 
aspects of general relativity in the laboratory.  
Indeed there are many works in the recent literature dealing with such
theme, not only in the context of light propagation in dielectric media; 
see for instance the review in Ref.\ \cite{Visser0}.
% and references therein. 

\acknowledgements
This work was partially supported by the Brazilian agencies 
{\em Conselho Nacional de Desenvolvimento 
Cient\'\i fico e Tecnol\'ogico} (CNPq) and 
{\em Funda\c c\~ao de Amparo \`a Pesquisa 
do Estado de Minas Gerais} (FAPEMIG).

\end{multicols}
\end{document}